\documentclass{easychairmod}
\usepackage{amsfonts,amsmath,amssymb,mathtools} 
\usepackage[only,llbracket,rrbracket,llparenthesis,rrparenthesis]{stmaryrd}
\usepackage{xcolor,listings}
\usepackage{tikz} 

\usepackage{doc}

\definecolor{keywordcolor}{rgb}{0.7, 0.1, 0.1}   
\definecolor{tacticcolor}{rgb}{0.0, 0.1, 0.6}    
\definecolor{commentcolor}{rgb}{0.4, 0.4, 0.4}   
\definecolor{symbolcolor}{rgb}{0.0, 0.1, 0.6}    
\definecolor{sortcolor}{rgb}{0.1, 0.5, 0.1}      
\definecolor{attributecolor}{rgb}{0.7, 0.1, 0.1} 

\definecolor{darkgreen}{rgb}{0.18,0.54,0.34}
\definecolor{darkpink}{rgb}{0.75,0.25,0.5}

\makeatletter
\lst@InputCatcodes
\def\lst@DefEC{%
	\lst@CCECUse \lst@ProcessLetter
	^^80^^81^^82^^83^^84^^85^^86^^87^^88^^89^^8a^^8b^^8c^^8d^^8e^^8f%
	^^90^^91^^92^^93^^94^^95^^96^^97^^98^^99^^9a^^9b^^9c^^9d^^9e^^9f%
	^^a0^^a1^^a2^^a3^^a4^^a5^^a6^^a7^^a8^^a9^^aa^^ab^^ac^^ad^^ae^^af%
	^^b0^^b1^^b2^^b3^^b4^^b5^^b6^^b7^^b8^^b9^^ba^^bb^^bc^^bd^^be^^bf%
	^^c0^^c1^^c2^^c3^^c4^^c5^^c6^^c7^^c8^^c9^^ca^^cb^^cc^^cd^^ce^^cf%
	^^d0^^d1^^d2^^d3^^d4^^d5^^d6^^d7^^d8^^d9^^da^^db^^dc^^dd^^de^^df%
	^^e0^^e1^^e2^^e3^^e4^^e5^^e6^^e7^^e8^^e9^^ea^^eb^^ec^^ed^^ee^^ef%
	^^f0^^f1^^f2^^f3^^f4^^f5^^f6^^f7^^f8^^f9^^fa^^fb^^fc^^fd^^fe^^ff%
	^^^^0393^^^^2081^^^^2082^^^^2208^^^^03c3
	^^00}
\lst@RestoreCatcodes
\makeatother

\lstdefinestyle{lean}{
	inputencoding=utf8,
	extendedchars=true,
	belowcaptionskip=1\baselineskip,
	breaklines=true,
	mathescape=true,
	keywordstyle=[1]{\ttfamily\color{keywordcolor}},
	keywordstyle=[2]{\ttfamily\color{sortcolor}},
	keywordstyle=[3]{\ttfamily\color{tacticcolor}},
	keywordstyle=[4]{\ttfamily\color{attributecolor}},
	commentstyle=\color{blue}\textit,
	stringstyle=\color{pink}\ttfamily,
	basicstyle=\footnotesize\ttfamily,
	showstringspaces=false,
	identifierstyle={\color{black}}, 
	literate=*
	{|}{{{\color{darkgreen}$|$}}}{1}
	{?}{{{\color{darkgreen}$\to$}}}{1}
	{:}{{{\color{darkgreen}:}}}{1}
	{c}{c}{1}
	{=}{{{\color{darkgreen}=\;}}}{1}
	{@}{{{\color{darkgreen}@}}}{1}
	{[}{{{\color{darkgreen}[}}}{1}
	{]}{{{\color{darkgreen}]}}}{1}
	{sigma}{{{\color{black}$\sigma$}}}{1}
	{.7}{{{\color{red}.7}}}{2},%
	morekeywords={
		import,open,reducible,noncomputable,notation,prefix,infix,local,attribute,
		example,definition,def,lemma,theorem,axiom,inductive,instance,structure,if,then,else,
		nat,bool,
		begin,end,by,match,with,cases,induction,fapply
	}
	morekeywords=[2]{Sort, Type, Prop},
	morekeywords=[3]{
		assumption,
		apply, intro, intros, allGoals,
		generalize, clear, revert, done, exact,
		refine, repeat, cases, rewrite, rw,
		simp, simp_all, contradiction,
		constructor, injection,
		induction,
	},
}

\usepackage{tikz}
\usetikzlibrary{arrows,calc,patterns,positioning,shapes}
\usetikzlibrary{decorations.pathmorphing}
\tikzset{
	modal/.style={>=stealth',shorten >=1pt,shorten <=1pt,auto,
		node distance=1.5cm,semithick},
	world/.style={circle,draw,minimum size=1cm,fill=gray!15},
	point/.style={circle,draw,fill=black,inner sep=0.5mm},
	reflexive/.style={->,in=120,out=60,loop,looseness=#1},
	reflexive/.default={5},
	reflexive point/.style={->,in=135,out=45,loop,looseness=#1},
	reflexive point/.default={25},
}

\usepackage{hyperref}

\usepackage{cleveref}

\title{Verified completeness in Henkin-style \\ for intuitionistic propositional logic}

\author{
Huayu Guo, Dongheng Chen, and Bruno Bentzen
}

\institute{
  School of Philosophy, Zhejiang University, Hangzhou, China \\
  \email{\{guohuayu,chen\_dongheng,bbentzen\}@zju.edu.cn} }

\authorrunning{H. Guo, D. Chen, and B. Bentzen}

\titlerunning{Verified completeness in Henkin-style for intuitionistic propositional logic}

\begin{document}

\maketitle

\begin{abstract}
 \noindent 
  This paper presents a formalization of the classical proof of completeness in Henkin-style 
  developed by Troelstra and van Dalen for intuitionistic logic with respect to Kripke models. 
  The completeness proof incorporates their insights in a fresh and elegant manner that is better suited for mechanization. 
  We discuss details of our implementation in the Lean theorem prover with emphasis on the prime extension lemma and construction of the canonical model. 
  Our implementation is restricted to a system of intuitionistic propositional logic with implication, conjunction, disjunction, and falsity given in terms of a Hilbert-style axiomatization. 
  As far as we know, our implementation is the first verified Henkin-style proof of completeness for intuitionistic logic following Troelstra and van Dalen's method in the literature. The full source code can be found online at \url{https://github.com/bbentzen/ipl}.
\end{abstract}

\section{Introduction}
\label{sect:introduction}

Troelstra and van Dalen~\cite{troelstravandalen1988constructivism} propose a completeness proof in Henkin-style for full intuitionistic predicate logic with respect to Kripke models. 
Despite being a fairly standard result in the literature, this completeness proof has yet to be formally verified in a proof assistant. 
In this paper, we describe a formalization for intuitionistic propositional logic using the Lean theorem prover~\cite{de2015lean}. 

Our main goal is to document some challenges encountered along the way and the design choices made to overcome them to obtain a formalized proof that is elegant, intuitive, and better suited for mechanization using the specific techniques available in the Lean programming language, in particular, the \texttt{encodable.decode} and \texttt{insert\_code} methods developed by Bentzen~\cite{bentzen2021henkin}. 




To the best of our knowledge, our implementation is the first verified Henkin-style proof of strong completeness for intuitionistic logic following Troelstra and van Dalen's method in the literature. 
As far as its propositional fragment is concerned, the main ingredient of Troelstra and van Dalen's Henkin-proof is a model construction based on a consistent extension of sets of formulas, which is achieved by going through all disjunctions of the language~\cite[lem 6.3]{troelstravandalen1988constructivism}. To carry out this extension, they assume an enumeration of disjunctions with infinite repetitions, also remarking that an alternative approach in which at each stage we treat the first disjunction not yet treated. This variant appears in Van Dalen~\cite[lem 5.3.8]{vandalen2013logic}. 
Our implementation is based on a third variant of the consistent extension method, which we developed to better suit our needs of formalization. Each propositional formula is only listed once in the enumeration, but we carry out the extension for each of them infinitely many times. 
The formalization consists of roughly 800 lines of code and encompasses the syntax and semantics of intuitionistic propositional logic, along with the soundness and strong completeness theorems. 
We adopt a Hilbert-style proof system due to its simplicity. The full source code can be found online at \url{https://github.com/bbentzen/ipl}.

\subsection{Related work}
\label{sect:related}

The formal verification of completeness proofs for intuitionistic logic can be traced back to Coquand's~\cite{coquand2002formalised} use of ALF to mechanize a constructive proof of soundness and completeness with respect to Kripke models for the simply typed lambda-calculus with explicit substitutions. 
Heberlin~and~Lee~\cite{herbelin2009forcing} give a constructive completeness proof of Kripke
semantics with constant domain for intuitionistic logic with implication
and universal quantification in Coq. Recently, Hagemeier and Kirst~\cite{hagemeier2022constructive}
formalize a constructive proof of completeness for intuitionistic epistemic logic based on a natural deduction system. 
They also provide a classical Henkin proof using methods similar to those in Bentzen~\cite{bentzen2021henkin}, but they do not present a formalization of the approach of Troelstra and van Dalen~\cite{troelstravandalen1988constructivism} as is done in this paper. 
Bentzen~\cite{bentzen2021henkin} formalizes the Henkin-style completeness method for modal logic S5 using Lean and From formalizes in Isabelle/HOL a Henkin-style completeness proof for both classical propositional logic~\cite{from2020formalizing} and classical first-order logic~\cite{from2022succinct}. 
Maggesi and Brogi~\cite{maggesi2021formal} give a formal completeness proof for provability logic in HOL Light. 
The formalization presented here is inspired by the work of Bentzen~\cite{bentzen2021henkin}, but makes a few improvements regarding design choices, in particular, the use of Prop in the definition of the semantics and the indexing of models to arbitrary types.

\subsection{Lean}
\label{sect:lean}

Lean~\cite{de2015lean} is an interactive theorem prover based on the version of dependent type theory known as the calculus of constructions with inductive types~\cite{pfenning1989inductively,coquand1986calculus}. Users can construct proof terms directly as in Agda~\cite{agda}, using tactics as in Coq~\cite{coq} or both proof terms and tactics simultaneously. Lean's built-in logic is constructive, but it supports classical reasoning as well. In fact, our Henkin-style proof is classical since it relies on a nonconstructive use of contraposition. 
Therefore, we do not worry about any complexity and computational aspects related to our proof. Our implementation makes use of some results from Lean's standard library and the user-maintained mathematical library \texttt{mathlib}~\cite{carneiro2018mathlib}. 

Throughout the remainder of this paper, Lean code will be used to showcase some design decisions in our formalization. 
The syntax and semantics of intuitionistic propositional logic that is the starting point of our formalization is described in~\Cref{sect:modal}. We also describe our formalization of a countermodel for the law of excluded middle and sketch a proof of soundness. Then, an informal overview of the Henkin-style proof method as well as a description of our implementation is provided in~\Cref{sect:completeness}. Finally, some concluding remarks are given in~\Cref{sec:conclusion}. 



\section{Intuitionistic Logic}
\label{sect:modal}

\subsection{The language}
\label{sect:modallogic}

The intuitionistic propositional language considered here contains implication, conjunction, disjunction, and falsity as the only primitive logical connectives. The language is defined using inductive types with one constructor for propositional letters, falsum, implication, conjunction, and disjunction, respectively:

\begin{lstlisting}[style=lean]
	inductive form : Type
	| atom : $\mathbb{N}$ $\rightarrow$ form
	| bot  : form
	| impl : form  $\rightarrow$ form  $\rightarrow$ form
	| and  : form  $\rightarrow$ form  $\rightarrow$ form
	| or   : form  $\rightarrow$ form  $\rightarrow$ form
\end{lstlisting}
This code can be found in \texttt{language.lean} file.

Since our language contains countably many propositional letters $p_0, p_1, ...$ we use the type $\mathbb{N}$ of natural numbers to define the constructor \texttt{atom} of propositional letters.
The only way to construct a term of type \texttt{form} is 
using this atomic constructor(\texttt{atom}) and the constructors for falsum (\texttt{bot}), implication (\texttt{impl}), conjunction (\texttt{and}), disjunction (\texttt{or}).

The elimination rule is an operation that allows us to define functions by recursion from it to any other types, including also the type of propositions \color{sortcolor}\texttt{Prop}\color{black}, in which case, this elimination rule is an instance of the principle of induction on the structure of the formula.

Constructors are displayed in Polish notation by default, but we define some custom infix notation with the usual Unicode characters for better readability:

\begin{lstlisting}[style=lean]
	prefix   `#`      := form.atom
	notation `$\bot$`       := form.bot
	infix    `$\supset$`       := form.impl
	notation p `&` q  := form.and p q
	notation p `$\vee$` q   := form.or p q
	notation `~`:40 p := form.impl p (form.bot )
\end{lstlisting}

\noindent Contexts are just sets of formulas.  In Lean sets are defined as functions of type \texttt{A} $\to$ \color{sortcolor}\texttt{Prop}\color{black}. As usual in logic textbooks, we display the formulas in a context in list notation separated by a comma instead of using unions of singletons. We introduce the following notation to make this possible:

\begin{lstlisting}[style=lean]
	notation $\Gamma$ ` $_{_{\grave{}}}$ ` p := set.insert p $\Gamma$ 
\end{lstlisting}

The formalization of the language can be found in the \texttt{language.lean} file.

\subsection{The proof system}
\label{sect:proofsystem}

We define a Hilbert-style system for intuitionistic propositional logic that is best described as a refinement of Heyting's original axiomatization~\cite[\S2]{heyting1930formalen}. The proof system is implemented with a type of proofs, which is inductively defined as follows: 

	
\begin{lstlisting}[style=lean]
	inductive prf : set form $\to$ form $\to$ Prop
	| ax {$\Gamma$} {p} (h : p $\in$ $\Gamma$) : prf $\Gamma$ p
	| k {$\Gamma$} {p q} : prf $\Gamma$ (p $\supset$ (q $\supset$ p))
	| s {$\Gamma$} {p q r} : prf $\Gamma$ ((p $\supset$ (q $\supset$ r)) $\supset$ ((p $\supset$ q) $\supset$ (p $\supset$ r)))
	| exf {$\Gamma$} {p} : prf $\Gamma$ ($\bot$ $\supset$ p)
	| mp {$\Gamma$} {p q} (hpq: prf $\Gamma$ (p $\supset$ q)) (hp : prf $\Gamma$ p) : prf $\Gamma$ q
	| pr1 {$\Gamma$} {p q} : prf $\Gamma$ ((p & q) $\supset$ p)
	| pr2 {$\Gamma$} {p q} : prf $\Gamma$ ((p & q) $\supset$ q)
	| pair {$\Gamma$} {p q} : prf $\Gamma$ (p $\supset$ (q $\supset$ (p & q)))
	| inr {$\Gamma$} {p q} : prf $\Gamma$ (p $\supset$ (p $\vee$ q))
	| inl {$\Gamma$} {p q} : prf $\Gamma$ (q $\supset$ (p $\vee$ q))
	| case {$\Gamma$} {p q r} : prf $\Gamma$ ((p $\supset$ r) $\supset$ ((q $\supset$ r) $\supset$ ((p $\vee$ q) $\supset$ r)))
\end{lstlisting}

Again, the elimination rule for this type generalizes definition by recursion and induction on the structure of proofs. To follow the usual logical notation, we abbreviate \texttt{prf $\Gamma$ p} with $\Gamma \vdash_i p$ as follows: 

\begin{lstlisting}[style=lean]
	notation $\Gamma$ ` $\vdash_i$ ` p := prf $\Gamma$  p
	notation $\Gamma$ ` $\not \vdash_i$ ` p := prf $\Gamma$  p $\to$ false
\end{lstlisting}

To illustrate, we compare a mechanized formal Hilbert-style proof of the identity of implication $p \supset p$ in our implementation:

\begin{lstlisting}[style=lean]
	lemma id {p : form } {$\Gamma$ :  set form } :
	| $\Gamma$ $\vdash_{i}$  p $\supset$ p :=
	mp (mp (@s  $\Gamma$ p (p $\supset$ p) p) k) k
\end{lstlisting}

\noindent with a non-mechanized formal proof written in Lemmon style:


\begin{center}
	\begin{tabular}{l | l l}
		1 \quad & \quad $ p \supset ((p \supset p)\supset p) \supset (p \supset (p \supset p)) \supset (p \supset p) $ & S\\
		2 & \quad $p \supset ((p\supset p)\supset p)$ & K\\
		3 & \quad $(p \supset (p \supset p)) \supset (p\supset p)$ & MP 1, 2\\
		4 & \quad $(p \supset (p \supset p))$ & K\\
		5 & \quad $p \supset p$ & MP 3, 4
	\end{tabular} 
\end{center}

Notice that the proof structure in our term proof is actually clearer since it indicates how the axiom schemes should be instantiated.

The formalization of the proof system can be found in the \texttt{theory.lean} file.

\subsection{Semantics}
\label{sect:semantics}

\subsubsection{Kripke models}
\label{sect:model}

We define the semantics for intuitionistic propositional logic in terms of Kripke semantics as usual~\cite{troelstravandalen1988constructivism,vandalen2013logic}. A model $\mathcal{M}$ is a triple $\langle \mathcal{W,\leq}, \mathsf{v} \rangle$ where $\mathcal{W}$ is a set of possible worlds of type $A$, $\leq$ is a reflexive, symmetric and monotonic binary relation on $A$, and $\mathsf{v}$ specifies the truth value of a formula at a world. 

In Lean, Kripke models can be defined as inductive types having just  one constructor using the \color{keywordcolor}\texttt{structure}\color{black}~command. We define it not as a triple but as a 6-tuple, composed of a domain~\texttt{W}, an accessibility relation~\texttt{R}, a valuation function~\texttt{val}, and proofs of reflexivity, transitivity, and monotonicity for the accessibility relation~\texttt{R}, denoted as \texttt{refl}, \texttt{trans}, and \texttt{mono}:

\begin{lstlisting}[style=lean]
	structure model (A : Type) :=
	| (W : set A)
	| (R : A  $\to$ A  $\to$ Prop)
	| (val : $\mathbb{N}$ $\to$ A $\to$ Prop)
	| (refl : $\forall$ w $\in$ W, R w w)
	| (trans : $\forall$ w $\in$ W, $\forall$ v $\in$ W, $\forall$ u $\in$ W, R w v $\to$ R v u $\to$ R w u)
	| (mono : $\forall$ p, $\forall$ w1 w2 $\in$ W, val p w1 $\to$ R w1 w2 $\to$  val p w2)
\end{lstlisting}


In our case, a possible world is a term of type $A$. This allows for more generality in the construction of a model unlike in \cite{bentzen2021henkin}. What is more, the type of propositions~\color{sortcolor}\texttt{Prop}\color{black}~is used to encode our truth values \texttt{true} or \texttt{false}.

\subsubsection{Semantic consequence}
\label{sect:forcing}

To formalize the notion of truth at a type, we define a forcing relation~$w \Vdash_{\mathcal{M}} p$ that takes as arguments a model~$\mathcal{M}$, a formula~$p$, and a type~$A$ and returns a term of type~\color{sortcolor}\texttt{Prop}\color{black}. As usual, falsity, conjunction, and disjunction are defined truth-functionally and an implication $ p \supset q$ is true at a world $w$ iff if $\mathcal{R}(w,v)$ then $p$ is true implies $q$ is true at $v$, for all $v \in \mathcal{W}$. We also introduce the familiar notation for this forcing relation:

\begin{lstlisting}[style=lean]
	def forces_form {A : Type} (M : model A) : form $\to$ A $\to$ Prop
	| (#p)     := $\lambda$v, M.val p v
	| (bot)    := $\lambda$v, false 
	| (p $\supset$ q)   := $\lambda$v, $\forall$ w $\in$ M.W, v $\in$ M.W $\to$ M.R v w 
	$\to$ forces_form p w $\to$ forces_form q w
	| (p & q)  := $\lambda$v, forces_form p v $\wedge$ forces_form q v
	| (p $\vee$ q)   := $\lambda$v, forces_form p v $\vee$ forces_form q v
	
	notation w `$\Vdash$  ` `{` M `} ` p := forces_form M p w
\end{lstlisting}

To formalize the intuitionistic notion of semantic consequence $\Gamma \vDash_i p$ we first extend this forcing relation to contexts pointwise and then we stipulate that $\Gamma \vDash_i p$ iff for all types~$A$, models~$\mathcal{M}$ and possible worlds~$w \in \mathcal{W}$, $\Gamma$ being true at $w$ in $\mathcal{M}$ implies $p$ being true at $w$ in $\mathcal{M}$: 

\begin{lstlisting}[style=lean]
	def forces_ctx {A : Type} (M : model A) ($\Gamma$ : set form) : A $\to$ Prop :=
	$\lambda$w, $\forall$ p, p $\in$ $\Gamma$ $\to$ forces_form M p w
	
	notation w `$\Vdash$` `{` M `} ` $\Gamma$ := forces_ctx M $\Gamma$ w
	
	def sem_csq ($\Gamma$ : set form) (p : form) := 
	$\forall$ {A : Type} (M : model A) (w $\in$ M.W), (w $\Vdash$ {M} $\Gamma$) $\to$ (w $\Vdash$ {M} p)
	
	notation $\Gamma$ `$\vDash_{i}$` p := sem_csq $\Gamma$ p
\end{lstlisting}

It is worth noting that we are overloading the forcing relation notation for formulas~\texttt{w $\Vdash$ \{M\} $p$} and contexts~\texttt{w $\Vdash$ \{M\} $\Gamma$}. There is no ambiguity because Lean will delay the choice until elaboration and determine how to disambiguate the notations depending on the relevant types.

The formalization of the Kripke semantics described above can be found in the \texttt{semantics.lean} file.

\subsubsection{The failure of the law of excluded middle}

Before proceeding to prove completeness, it will be helpful to see how we can build models in our implementation. To give a concrete example, let us show how to build the following countermodel for the law of excluded middle~\cite[p.99]{kripke1965semantical} using the type of booleans true~$\mathtt{tt}$ and false~$\mathtt{ff}$:

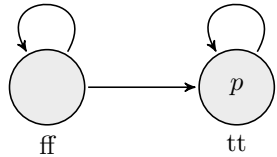
\begin{center}
	\begin{tikzpicture}[modal]
		\node[world] (w) [label=below:ff] {};
		\node[world] (v) [label=below:tt,right=of w] {$p$};
		\path[->] (w) edge (v);
		\path[->] (w) edge[reflexive] (w);
		\path[->] (v) edge[reflexive] (v);
	\end{tikzpicture}
\end{center}

Since our possible worlds are always booleans, the domain, accessibility relation, and valuation function are formalized in Lean in a slightly different way. The reflexivity, transitivity, and monotonicity proofs are straightforward, so we shall omit them:


\begin{lstlisting}[style=lean]
	def W : set bool := {ff, tt}
	
	def R : bool $\rightarrow$ bool $\rightarrow$ Prop := $\lambda$ w v, w = v $\lor$ w = ff
	
	@[simp]
	def val : nat $\rightarrow$ bool $\rightarrow$ Prop := $\lambda$ _ w, w = tt
	
\end{lstlisting}

Using this countermodel, we assume that the law of excluded middle holds, that is for any formula $p$, either $\emptyset \models_i p$ or $\emptyset \models_i \neg p$, and then derive a contradiction. This allows us to prove that the law of excluded middle fails in general:

\begin{lstlisting}[style=lean]
	lemma no_lem: $\neg$ $\forall$ p, ($\emptyset$ $\vDash_i$  p $\lor$ ~p) 
\end{lstlisting}

The mechanization of the countermodel can be found in the \texttt{nolem.lean} file.

\subsubsection{Soundness}

The soundness theorem asserts that if a formula $p$ can be derived from a set of assumptions $\Gamma$ using the inference rules of the logical system, then $p$ is logically valid under any interpretation that satisfies $\Gamma$.

\begin{lstlisting}[style=lean]
	theorem soundness {$\Gamma$ :  set form} {p : form} : 
	($\Gamma$ $\vdash_{i}$ p) $\rightarrow$ ($\Gamma$ $\models_i$  p)
\end{lstlisting}

The code for proof of soundness can be found in \texttt{soundness.lean}.

The proof proceeds by using induction to perform case analysis for each inference rule. For each rule, the proof provides a way to derive the conclusion based on the rule and a way to show that the conclusion is logically valid based on the interpretation and the premises. 

\section{The completeness theorem} 
\label{sect:completeness}


Now that we have presented the implementation of the syntax and semantics of intuitionistic propositional logic in the previous section, we are prepared to undertake a formal proof of completeness. 
The strong completeness theorem, which states that every semantic consequence is a syntactic consequence, can be stated in Lean using our custom notation as follows:

\begin{lstlisting}[style=lean]
	theorem completeness {$\Gamma$ :  set form} {p : form} : 
	($\Gamma$ $\models_i$  p) $\rightarrow$ ($\Gamma$ $\vdash_{i}$ p) 
\end{lstlisting}

Our implementation follows the original Henkin-style completeness proof given by Troelstra and van Dalen~\cite{troelstravandalen1988constructivism} with some small modifications.
The main proof argument runs as follows. 

\begin{enumerate}
	\item Assume that $\Gamma \vDash_{i} p$ and  $\Gamma \nvdash_{i} p$ hold;
	\item Build a model $\mathcal{M}$ such that $w \Vdash_\mathcal{M} p$ iff $w \vdash_i p$ for all worlds $w \in \mathcal{W}$, where we have sets of formulas as possible worlds;
	
	\item Show that there is a world $w \in \mathcal{W}$ such that $w \Vdash_{\mathcal{M}} \Gamma$ but  $w \nVdash_{\mathcal{M}} p$;
	\item Establish a contradiction from our assumption that $\Gamma \vDash_{i} p$.  
\end{enumerate}

Our proof appeals to classical reasoning at the metalevel of Lean's logic on two occasions~\cite[p.87]{troelstravandalen1988constructivism}, namely,  
in our proof of $\Gamma \vdash_{i} p$ where we assume double negation elimination and in our proof of $w \Vdash_\mathcal{M} p$ iff $w \vdash_i p$.

The reader can refer to the \texttt{completeness.lean} file for the full details of our implementation of the completeness proof.	

\subsubsection{Consistent prime extensions}
\label{sect:extension}
The first step of Troelstra and van Dalen's  proof is the definition of what they call a ``saturated theory''~\cite[def.6.2]{troelstravandalen1988constructivism}. We shall make use of the equivalent concept of prime theory instead~\cite[def.5.3.7]{vandalen2013logic}, in which the disjunction property is expressed in terms of the membership relation.
We say that a set of formulas $\Gamma$ is a prime theory if $\Gamma$ is closed under derivability and if $p \lor q \in \Gamma$ implies $p \in \Gamma$ or $q \in \Gamma$. In \texttt{completeness.lean} file, we write:

\begin{lstlisting}[style=lean]
	def is_closed ($\Gamma$ :  set form) :=  
	$\forall$ {p : form}, ($\Gamma$ $\vdash_{i}$ p) $\rightarrow$ p $\in$ $\Gamma$
	
	def has_disj ($\Gamma$ :  set form) := 
	$\forall$ {p q : form}, ((p $\lor$ q) $\in$ $\Gamma$) $\rightarrow$ ((p $\in$ $\Gamma$) $\vee$ (q $\in$ $\Gamma$))
	
	def is_prime ($\Gamma$  : set form) := 
	is_consist $\Gamma$ $\land$ has_disj $\Gamma$ 
\end{lstlisting}
The second step of Troelstra and van Dalen's completeness proof is the proof of a prime extension lemma~\cite[lem 6.3]{troelstravandalen1988constructivism}, which states that if $\Gamma \nvdash r$ then there is a prime theory $\Gamma' \supseteq \Gamma$ such that $\Gamma' \nvdash r$.
Assuming that they have a list of disjunctions $\langle \varphi_{i,1} \lor \varphi_{i,2} \rangle_i$ with infinite repetitions, 
they define $$\Gamma' = \bigcup_{i \in \mathbb{N}} \Gamma_i,$$ where $ \Gamma_0= \Gamma$ and $ \Gamma_{k+1}$ is defined inductively as follows:

\begin{itemize}
	\item Case 1: $\Gamma_k \vdash \varphi_{k,1} \lor \varphi_{k,2}$. Put \\
	\begin{itemize}
		\item $\Gamma_{k+1}=\Gamma_k \cup \{\varphi_{k,2}\}$ if $\Gamma_k, \varphi_{k,1} \vdash r$, and \\
		\item $\Gamma_{k+1}=\Gamma_k \cup \{\varphi_{k,1}\}$ otherwise \\
	\end{itemize}
	
	\item Case 2: $\Gamma_k \nvdash \varphi_{k,1} \lor \varphi_{k,2}$. Put \\
	\begin{itemize}
		\item $\Gamma_{k+1}=\Gamma_k $
	\end{itemize}
	
\end{itemize}

Since we want to extend $\Gamma$ to a prime theory $\Gamma'$, we want to ensure the disjunctive property that if $\phi \lor \psi \in \Gamma'$ then $\phi \in \Gamma'$ or $\psi \in \Gamma'$. 
If there were no infinite repetitions in the list, we could never be sure that we have treated all disjunctions in Case 1, for, at step $k+1$, its disjuncts only get added to the set when $\Gamma_k$ proves the disjunction. It is possible that later the disjunction becomes provable from $\Gamma_{k+m}$, but, we will never go back to it again. 

Troelstra and van Dalen mention a simpler variant of the construction that uses an enumeration of disjunctions without requiring infinite repetitions. At stage $k+1$ we simply treat the first disjunction not yet treated. This proof is spelled out by van Dalen in~\cite[lem 5.3.8]{vandalen2013logic}. However, the proof method is less suitable for mechanization given that it is difficult to tell a proof assistant how exactly they should find the first disjunction not yet treated. 
We implement a simplified version of this method where at each step $k+1$ we always treat all disjunctions in the language once more. 
The following Lean code encapsulates the idea of the construction sketched above: 

\begin{lstlisting}[style=lean]
	def insert_form ($\Gamma$ :  set form) (p q r : form) :  set form :=  
	if ($\Gamma$ $_{_{\grave{}}}$ p $\vdash_{i}$ r) then $\Gamma$ $_{_{\grave{}}}$ q else $\Gamma$ $_{_{\grave{}}}$ p
	
	def insert_code ($\Gamma$ :  set form) (r : form) (n : nat) :  set form :=
	match encodable.decode (form) n with
	| none   := $\Gamma$
	| some (p $\lor$ q) := if $\Gamma$ $\vdash_{i}$ p $\lor$ q then insert_form $\Gamma$ p q r else $\Gamma$
	| some _ := $\Gamma$
	end
	
	def insertn ($\Gamma$ :  set form) (r : form) : nat $\rightarrow$  set form
	| 0     := $\Gamma$ 
	| (n+$1$) := insert_code (insertn n) r n 
	
	def primen ($\Gamma$ :  set form) (r : form) : nat $\rightarrow$  set form
	| 0     := $\Gamma$
	| (n+$1$) := $\bigcup$ i, insertn (primen n) r i 
	
	def prime ($\Gamma$:  set form) (r : form) :  set form :=
	$\bigcup$ n, primen $\Gamma$ r n
\end{lstlisting}

Unlike in Troesltra and van Dalen~\cite{troelstravandalen1988constructivism} and van Dalen~\cite{vandalen2013logic}, the enumeration in our formalization lists not just all disjunctions but all propositional formulas in the language. When a formula is not a disjunction we simply ignore it just as in Case~2 above. We follow Bentzen~\cite{bentzen2021henkin} in using \texttt{encodable} types to enumerate the language. 
In Lean, a type $\alpha$ is encodable if there is an encoding function  \lstinline[style=lean]{encode : $\alpha$ $\to$ nat} and a (partial) inverse \lstinline[style=lean]{decode : nat $\to$ option $\alpha$} that decodes the encoded term of $\alpha$.

Now that we extended $\Gamma$ to $\Gamma'$, which we denote as \texttt{prime $\Gamma$ r}, we have to prove it is indeed a prime extension of $\Gamma$.
First, we show that $\Gamma \subseteq \Gamma'$. But this is easy, since for every $\Gamma_n'$ \texttt{n} in the family of sets, $\Gamma \subseteq \Gamma_n' \texttt{ n}$.  Therefore, $\Gamma$ must also be included in the union of all $\Gamma_n'$ \texttt{n}, which is $\Gamma_n'$.

\begin{lstlisting}[style=lean]
	lemma primen_subset_prime {$\Gamma$ :  set form} {r : form} (n): 
	primen $\Gamma$ r n $\subseteq$  prime $\Gamma$ r
	
	lemma subset_prime_self {$\Gamma$ :  set form} {r : form} :
	$\Gamma$ $\subseteq$ prime $\Gamma$  r 
	
\end{lstlisting}

The next step is to prove that the $\Gamma'$ also has the disjunction property and it is closed under derivability. Let us focus on the former first. 

We need to show that $p \lor q \in \Gamma'$ implies $p \in \Gamma'$ or $q \in \Gamma'$. If $p \lor q \in \Gamma'$ then there is some $n \in \mathbb{N}$ such that $p \lor q \in \Gamma_n'$. But then since $\Gamma_n' \vdash p \lor q$, then we know that $p \in \Gamma_{n+1}'$ or $q \in \Gamma_{n+1}'$ because the disjunction was treated at some point. Thus, $p \in \Gamma'$ or $q \in \Gamma'$.

\begin{lstlisting}[style=lean]
	def prime_insertn_disj {$\Gamma$:  set form} {p q r : form} (h : (p $\lor$ q)   $\in$ prime $\Gamma$ r) : 
	$\exists$ n, p $\in$ (insertn (primen $\Gamma$ r n) r (encodable.encode (p $\lor$ q)+$1$)) $\lor$ q $\in$ (insertn (primen $\Gamma$ r n) r (encodable.encode (p $\lor$ q)+$1$))
	
	lemma insertn_to_prime {$\Gamma$ :  set form} {r : form} {n m : nat} : 
	insertn (primen $\Gamma$ r n) r m $\subseteq$ prime $\Gamma$  r 
	
	def prime_has_disj {$\Gamma$ :  set form} {p q r : form} : 
	((p $\lor$ q) $\in$ prime $\Gamma$   r) $\rightarrow$ p $\in$ prime $\Gamma$   r $\vee$ q $\in$ prime $\Gamma$  r 
\end{lstlisting}

Saying that $\Gamma'$ is closed under derivability means that if we can deduce a formula from $\Gamma'$, it is an element of $\Gamma'$. We use a lemma that states that if we can prove $r \lor p$ from $\Gamma'$, then there exists an $n$ such that $p \in \Gamma_{n+1}$. We use the above lemma \texttt{insertn\_to\_prime} to deduce that $p \in \Gamma'$:

\begin{lstlisting}[style=lean]
	lemma prime_prf_disj_self {$\Gamma$ :  set form} {p r : form} :  
	(prime $\Gamma$ r $\vdash_{i}$   r $\lor$ p) $\rightarrow$ $\exists$ n, p $\in$ (insertn (primen $\Gamma$ r n) r (encodable.encode (r $\lor$ p)+$1$)) 
\end{lstlisting}
\begin{lstlisting}[style=lean]
	def prime_is_closed {$\Gamma$ :  set form} {p q r : form} : 
	(prime $\Gamma$ r $\vdash_{i}$ p) $\rightarrow$ p $\in$ prime $\Gamma$  r
\end{lstlisting}

At this moment, we need to prove that $\Gamma'$ still remains consistent. First, we by structural induction on the derivation that if $\Gamma' \vdash r$ then there is some $n$ such that $\Gamma_n \vdash r$. Then we prove by induction on $n$ that if $\Gamma_n \vdash r$ then $\Gamma \vdash r$. The base case is trivial. In the inductive case, we complete the proof by unfolding the definition of $\Gamma_n$ and manipulating the inductive hypothesis. Putting both lemmas together, we prove that $\Gamma' \vdash r$ implies $\Gamma \vdash r$:


\begin{lstlisting}[style=lean]
	def primen_not_prfn {$\Gamma$ :  set form} {r : form} {n} : 
	(primen $\Gamma$ r n $\vdash_{i}$ r) $\rightarrow$ ($\Gamma$ $\vdash_{i}$ r)
	
	def prime_not_prf {$\Gamma$ :  set form} {r : form} : 
	(prime $\Gamma$ r $\vdash_{i}$ r) $\rightarrow$ ($\Gamma$ $\vdash_{i}$ r)
\end{lstlisting}


\subsubsection{The canonical model construction}
\label{sect:canonicalmodel}

Given a set of formulas~$\Gamma$ and $\phi$ such that $\Gamma \nvdash \phi$, the next step is to build a canonical Kripke model $\mathcal{M}$ such that with $w \Vdash_\mathcal{M} \Gamma$ and $w \nVdash_\mathcal{M} \phi$ for some possible world.
We build this model by letting $\mathcal{W}$ be the set of all consistent prime theories; $w \leq v$ iff $w \subseteq v$ for $w,v \in \mathcal{W}$; and $\textsf{v}(w,p)=1$ iff $w\in \mathcal{W} $ and $p \in w$, for a propositional letter $p$.
The following Lean code reflects the model construction: 

\begin{lstlisting}[style=lean]
	def domain : set (set form) := {w | is_consist w $\land$  ctx.is_prime w}
	
	def access : set form $\rightarrow$ set form $\rightarrow$ Prop := $\lambda$ w v, w $\subseteq$ v
	
	def val : $\mathbb{N}$ $\rightarrow$ set form $\rightarrow$ Prop := $\lambda$ q w, w $\in$ domain $\land$ (#q) $\in$ w
	
\end{lstlisting}




\noindent The accessibility relation $\leq$ is clearly reflexive and transitive since so is $\subseteq$. Monotonicity is easy to see since $p \in w$ and $w \subseteq v$ means that $q \in v$. We prove these lemmas by  straightforward unfolding the definition of \texttt{access}.

Our model is integrated into Lean's code as follows:

\begin{lstlisting}[style=lean]
	def M : model (set form):=
	begin
	fapply model.mk,
	apply domain,
	apply access,
	apply val,
	apply access.refl,
	apply access.trans,
	apply access.mono
	end
\end{lstlisting}
\subsubsection{Truth and derivability}
\label{sect:fullproof}

It turns out that a formula is true at a world in the canonical model if and only if it can be proved from that world:

\begin{lstlisting}[style=lean]
	lemma model_tt_iff_prf {p : form} : 
	$\forall$ (w $\in$ domain), (w $\models$ {M} p) $ \leftrightarrow$ (w $\vdash_{i}$ p) 
\end{lstlisting}

\noindent We mechanize the proof employing the \texttt{\color{tacticcolor}induction} tactic, which allows us to use the elimination rule of a type. This approach yields five goals, namely, to prove the case where a formula is a propositional letter, falsity, implication, conjunction, or disjunction. The proof of implication and disjunction deserve some mention. 

The disjunction case is simpler, so we shall discuss it first. Lean gives us a biconditional in the following goal: 

\begin{lstlisting}[style=lean]
	$\vdash$ $\forall$ (w : set form),
	w $\in$ domain $\rightarrow$ (w $\models$ {M} (p $\lor$ q))  $\leftrightarrow$ (w $\vdash_{i}$  p $\lor$ q))
\end{lstlisting}

The proof in the forward direction starts with the introduction of assumptions and then splits the proof into two cases.
In the first case, we assume that $w \models_{\mathcal{M}} p \lor q$ and our goal is $w \vdash_{i} p \lor q$. Through the tactic \texttt{cases}, which expresses case reasoning, we can finish our goal using some basic facts about disjunctions and the inductive hypotheses in both cases.

In the backward direction, we assume that $w \vdash_{i} p \lor q$. Since $w$ is a prime theory and thus enjoys the disjunctive property, we can reason by cases depending on whether $w \vdash_{i} p$ or $w \vdash_{i} q$. The result follows the inductive hypothesis.


Now we proceed to the implication case. Using the \texttt{intro} tactic, we begin by assuming the inductive hypothesis for $p$. If $w$ is a world and it is a prime theory, then by unfolding the true definition of a formula in the model's world, we arrive at a biconditional goal that can be expressed as follows.

\begin{lstlisting}[style=lean]
	$\vdash$ $\forall$ (w : set form),
	w $\in$ domain $\rightarrow$ (w $\models_i$ {M} (p $\supset$ q)) $\leftrightarrow$ (w $\vdash_{i}$ p $\supset$ q))
\end{lstlisting}

We split the biconditional proof into two smaller conditionals using the \texttt{split} tactic.  
In the forward direction, we first assume that $w \Vdash_{\mathcal{M}} p \supset q$. We reason by cases depending on whether $w \vdash_i p \supset q$ or not, therefore invoking the law of excluded middle. If that is the case, we are done. If not, then we know that $w, p \nvdash q$. We want to derive a contradiction. We extend the context $w, p$ to a prime theory $(w, p)'$ that still does not prove $q$. 
By our inductive hypothesis, since $(w, p)'$ is in the domain, we know that $(w,p)' \Vdash_{\mathcal{M}} q \leftrightarrow (w,p)' \vdash_i q$. 

To derive a contradiction, we just have to show that $(w,p)' \Vdash_{\mathcal{M}} q$. Recall that our assumption $w \Vdash_{\mathcal{M}} p \supset q$ states that for all $v \in \mathcal{W}$ such that $w \leq v$, if $v \Vdash_{\mathcal{M}} p$ then $v \Vdash_{\mathcal{M}} q$. But, clearly, $w \leq (w,p)'$. To complete the proof, we just have to show that $(w,p)' \Vdash_{\mathcal{M}} p$. By our inductive hypothesis, it suffices to show that $(w,p)' \vdash_i p$. But this is clearly true, since the original set $w,p$ is contained in the prime extension $(w,p)'$ and $w,p \vdash_i p$. 



For the backward direction,  what we have to prove is $ w \Vdash_{\mathcal{M}} p \supset q$. This means for all $v \in \mathcal{W}$ such that $w \leq v$,  if $v \Vdash_{\mathcal{M}} p$ then $v \Vdash_{\mathcal{M}} q$. We assume that $v \in \mathcal{W}$ such that $w \leq v$, $v \Vdash_{\mathcal{M}} p$ then  we have to show $ v \Vdash_{\mathcal{M}} q$. Using our inductive hypothesis, we just have to show that $v \vdash_i  q$.

Since we know $w \vdash_i p \supset q$ and $w \subseteq v$, by weakening, we will have $v \vdash_i p \supset q$. We complete the proof by noting that $v \vdash_i p$ by our inductive hypothesis and assumption that $v \Vdash_{\mathcal{M}} p$. The result follows from modus ponens.


We have finished the proof of implication.

\subsubsection{The completeness proof}
\label{sect:fullproof}

To finish our completeness proof we just have to put together all the above pieces into 27 lines of code. We assume that $\Gamma \nvdash_{i} p$ and $\Gamma \models_i p$, we just need to arrive at a contradiction. 
We extend $\Gamma$ to a prime theory $\Gamma'$ such that $\Gamma' \nvdash_{i} p$. Since we know $\Gamma' \Vdash_{\mathcal{M}} q \iff \Gamma' \vdash_{i} q$ for every formula $q$, we can conclude that $\Gamma' \nVdash_{\mathcal{M}} p$. Thus, we contradict our assumption that $\Gamma \models_i p$, given that $\Gamma' \Vdash_{\mathcal{M}} \Gamma$ but $\Gamma' \nVdash_{\mathcal{M}} p$. 

\section{Conclusion}
\label{sec:conclusion}

We have used Lean to formally verify the Henkin-style completeness proof for intuitionistic logic proposed by Troesltra and van Dalen~\cite{troelstravandalen1988constructivism} restricted to a propositional fragment with implication, falsity, conjunction, disjunction. The propositional proof system we implement is based on a Hilbert-style axiomatization. In future work, we hope to expand our implementation to full intuitionistic first-order logic with existential and universal quantifiers and thus complete the formalization of Troesltra and van Dalen's proof. Our implementation also includes a mechanized proof of soundness and a countermodel for the general validity of the law of excluded middle in intuitionistic propositional logic.\\

\vspace{5mm}
\noindent \textbf{Acknowledgments} \;  
This research was supported in part by the Zhejiang Federation of Humanities and Social Sciences grant 23YJRC04ZD.

\label{sect:bib}


\end{document}